\documentclass[a4paper,11pt]{article}
\usepackage{pos}
\usepackage{siunitx}
\usepackage{subcaption}
\usepackage{hyperref}
\usepackage{lineno}

\DeclareSIUnit \h {\ensuremath{\mathit{h}}}
\DeclareSIUnit \pc {pc}
\DeclareSIUnit \Mpc {Mpc}
\DeclareSIUnit \kpc {kpc}
\DeclareSIUnit \erg {erg}
\DeclareSIUnit \deg {deg}
\DeclareSIUnit \year {yr}
\DeclareSIUnit \Gyear {Gyr}
\DeclareSIUnit \gauss {G}
\DeclareSIUnit \EeV {EeV}
\DeclareSIUnit \PeV {PeV}
\DeclareSIUnit \tons {t}
\DeclareSIUnit \sr {sr}

\title{Application of pattern spectra and convolutional neural networks to the analysis of simulated Cherenkov Telescope Array data}
 \ShortTitle{Application of pattern spectra and CNNs on CTA data}

\author*[a]{J. Aschersleben}
\author[a]{R. F. Peletier}
\author[a]{M. Vecchi}
\author[b]{M. H. F. Wilkinson}

\affiliation[a]{Kapteyn Astronomical Institute\\
University of Groningen, PO Box 800, 9700 AV Groningen, The Netherlands}

\affiliation[b]{Bernoulli Institute for Mathematics, Computer Science and Artificial Intelligence\\
University of Groningen, PO Box 407, NL-9700 AK Groningen, The Netherlands}

\forColl{CTA} 

\emailAdd{j.j.m.aschersleben@rug.nl}
\emailAdd{r.f.peletier@rug.nl}
\emailAdd{m.vecchi@rug.nl}
\emailAdd{m.h.f.wilkinson@rug.nl}

\abstract{The Cherenkov Telescope Array (CTA) will be the next generation gamma-ray observatory and will be the major global instrument for very-high-energy astronomy over the next decade, offering $5 - 10$ $\times$ better flux sensitivity than current generation gamma-ray telescopes. Each telescope will provide a snapshot of gamma-ray induced particle showers by capturing the induced Cherenkov emission at ground level. The simulation of such events provides images that can be used as training data for convolutional neural networks (CNNs) to determine the energy of the initial gamma rays. Compared to other state-of-the-art algorithms, analyses based on CNNs promise to further enhance the performance to be achieved by CTA. \\
Pattern spectra are commonly used tools for image classification and provide the distributions of the shapes and sizes of various objects comprising an image. The use of relatively shallow CNNs on pattern spectra would automatically select relevant combinations of features within an image, taking advantage of the 2D nature of pattern spectra. In this work, we generate pattern spectra from simulated gamma-ray events instead of using the raw images themselves in order to train our CNN for energy reconstruction. This is different from other relevant learning and feature selection methods that have been tried in the past. Thereby, we aim to obtain a significantly faster and less computationally intensive algorithm, with minimal loss of performance.}

\FullConference{37$^{\rm{th}}$ International Cosmic Ray Conference (ICRC 2021)\\
		July 12th -- 23rd, 2021\\
		Online -- Berlin, Germany}

\begin{document}
\maketitle

\section{Introduction}
\noindent
The interaction of a gamma ray with the Earth atmosphere induces a particle shower, which produces a flash of Cherenkov light. Imaging atmospheric Cherenkov telescopes (IACTs) can capture the Cherenkov emission at ground level, which enables the estimation of the energy of the initial gamma ray. The Cherenkov Telescope Array (CTA)\footnote{www.cta-observatory.org} will host the next generation of IACTs and will offer a $5 - 10$ $\times$ better flux sensitivity than current generation gamma-ray telescopes~\cite{CTA2018}. The telescopes will be located in both the northern and southern hemispheres at the Roque de los Muchachos Observatory in La Palma (CTA North) and the Atacama Desert in Chile (CTA South). The combination of telescopes of three different sizes: Small-Sized Telescopes (SSTs), Medium-Sized Telescopes (MSTs) and Large-Sized Telescope (LSTs) will provide a wide energy range between $\SI{20}{\GeV}$ and $\SI{300}{\TeV}$ and a precision of $\sim \SI{1}{\arcmin}$ on individual photons for the upper end of the CTA energy range, which is the best resolution achieved anywhere above the X-ray domain. 

Convolutional neural networks (CNNs) are a subclass of artificial neural networks (ANNs)~\cite{li2020survey} and can be trained with the Cherenkov images simulated for CTA to determine the energy of the initial gamma rays. Compared to other state-of-the-art algorithms, analyses based on CNNs promise to further improve the performance to be achieved by CTA~\cite{Jacquemont2021,Jacquemont2021_2,Mangano2018,Nieto2021}. However, the construction of CNNs is typically very computationally expensive due to a large number of free parameters and the amount of data required. Pattern spectra~\cite{192465} are commonly used tools for image classification, which provide the distributions of the shapes and sizes of various objects comprising an image and can significantly reduce the computational power needed to train a CNN. They are constructed using a technique from mathematical morphology known as granulometries~\cite{BREEN1996377}, which can be computed with connected operators~\cite{salmbier2009}. Compared to other classical approaches, connected operators have the advantage of not introducing any distortions into the image (see Ref.~\cite{Nieto2019} for a detailed study of image distortion in IACT event reconstruction with neural networks). This is achieved by the merging of flat zones (regions in the image with the same colour) within the image, which prevents splitting or deforming of existing features and the implementation of unwanted new edges. 

\section{Dataset}
\begin{figure}[h]
    \centering
    \includegraphics[width=0.6\textwidth]{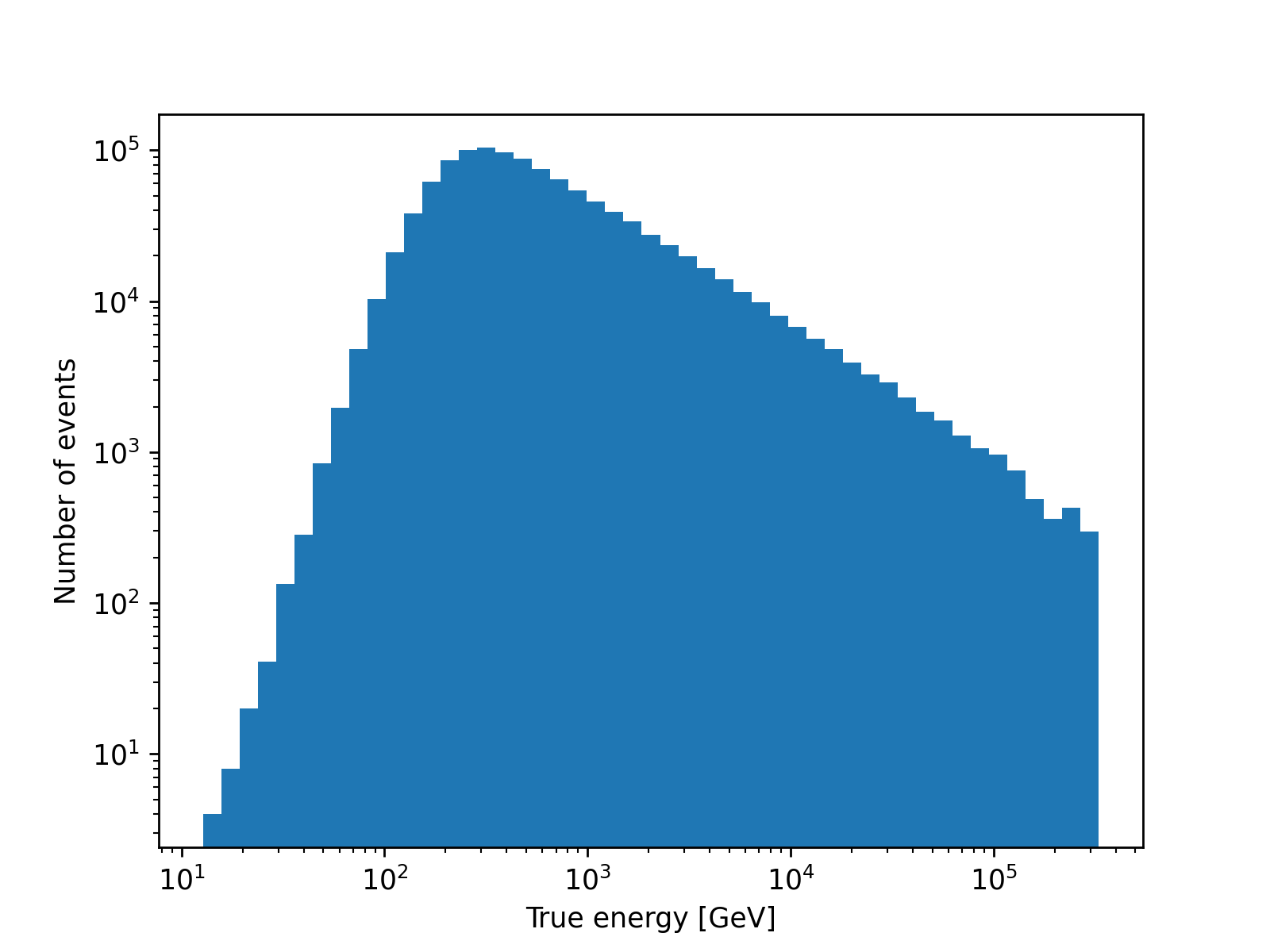}
    \caption{Energy distribution of the simulated gamma-ray events}
    \label{fig:dataset}
\end{figure}

\noindent
The dataset consists of simulated shower images of gamma-ray events with CTA South (zenith angle of $\SI{20}{\degree}$, North pointing) generated with a $\SI{0.4}{\degree}$ offset from the telescope pointing position. In this analysis only the charge information (i.e. the integrated photodetector pulse) of SST images is considered.
The energy distribution of the events is shown in Figure~\ref{fig:dataset} and covers an energy range between $\SI{20}{\GeV}$ and $\SI{300}{\TeV}$. The energy used to simulate each individual event is referred to $E_{\text{true}}$ in the following.\\
In order to achieve the best CNN performance a large number of gamma-ray events is required. Therefore, the dataset consists of $\sim 1 \times 10^{6}$ gamma-ray events. Depending on the initial gamma-ray energy and the impact position and direction, a single event can be captured by several SSTs. As a first step towards the implementation of pattern spectra for the analysis of CTA images, images of the same event captured by several SSTs are combined into one single image by adding up the individual pixel values of each image.

\section{Analysis}
\begin{figure}[h]
    \centering
    \includegraphics[width=\textwidth,trim={0cm 4cm 0cm 4cm},clip]{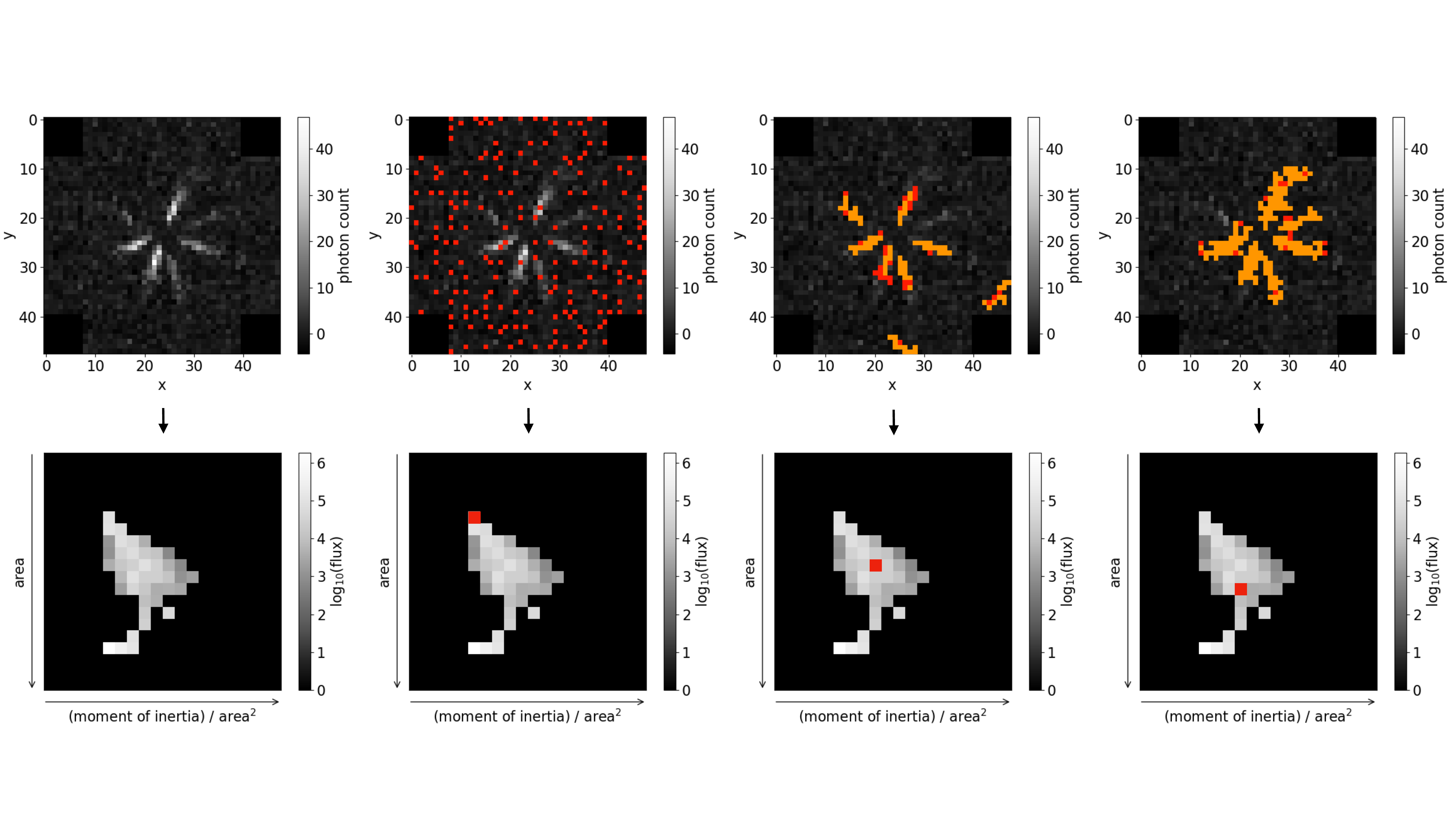}
    \caption{Top: 8-bit CTA images with highlighted features (in red/orange) detected by the pattern spectra software. Bottom: pattern spectra with the pixel (in red) corresponding to the detected features.}
    \label{fig:ps_examples}
\end{figure}

\noindent
The pattern spectra algorithm is based on the algorithm presented in Urbach et al.(2007)~\cite{Urbach2007}, which creates 2-dimensional (size $\&$ shape) pattern spectra. 
It detects objects within the image, which are the connected components of threshold sets of the image.
The size of the objects in the image is classified by the area of the object $A$. The shape of the objects in the image is classified by $I/A^2$, which is the ratio of the moment of inertia $I$ to the square of the area $A$. The moment of inertia $I$ describes the sum of squared differences to the centre of gravity of the object.
For a more detailed definition of the area $A$ and the moment of inertia $I$, see Urbach et al. (2007)~\cite{Urbach2007}. In order to construct a pattern spectrum from a CTA image, the CTA image has to be converted into an 8-bit greyscale image in PGM format due to current software limitations. This conversion includes a loss of information, which will be discussed in more detail in the last section. An example of a pattern spectrum obtained from a $\sim \SI{1.9}{\TeV}$ gamma-ray event image is shown in Figure~\ref{fig:ps_examples}. The top-left image shows the input image and the bottom-left image the corresponding pattern spectrum. The remaining three image pairs show the detected features in the input image (features highlighted in orange, subfeatures highlighted in red) corresponding to the specific pixel in the pattern spectrum (marked in red). Whereas the features detected in the second image correspond mostly to noise, the features in the third and fourth image correspond to the Cherenkov photons emitted by the particle shower, which are of particular interest for energy reconstruction.

\begin{figure}[h]
    \centering
    \includegraphics[width=0.8\textwidth,trim={17cm 10cm 17cm 10cm},clip]{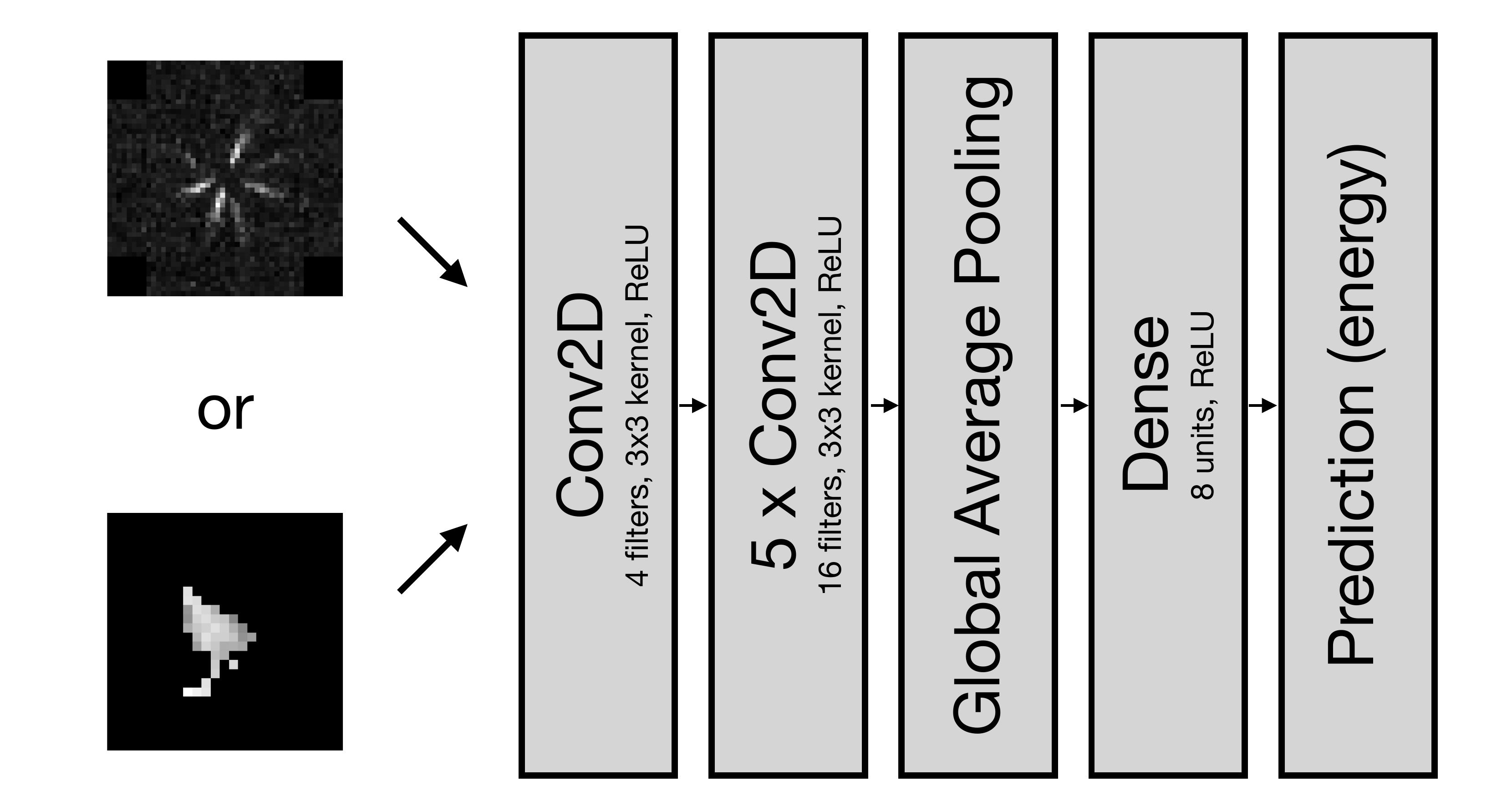}
    \caption{Sketch of the CNN architecture used for this analysis. Although the CTA images and the pattern spectra are applied separately on the CNN, the CNN architecture is the same in both cases.}
    \label{fig:cnn}
\end{figure}

Taking either the original CTA images, 8-bit CTA images or pattern spectra as input, the CNN provides the (reconstructed) energy as an output, which is referred to $E_{\text{rec}}$ in the following. The 8-bit CTA image analysis operates as a reference in order to get a rough estimate for the loss of information during the 8-bit conversion.
The CNN is constructed using \textit{Tensorflow 2.3.1}~\cite{Tensorflow_bib} and \textit{Keras 2.4.3}~\cite{Keras_bib}. It consists of six convolutional layers, followed by a global average pooling and a dense layer, which results in a total of 26,729 free parameters. For training the CNN, a batch size of 32, the ADAptive Moment (ADAM) optimizer~\cite{Kingma2017}, a constant learning rate of $10^{-3}$, the mean squared error as loss function and 50 epochs were chosen. The dataset was split into $\SI{90}{\percent}$ training data (of which $\SI{10}{\percent}$ was used for validation) and $\SI{10}{\percent}$ test data. An illustration of our CNN is shown in Figure~\ref{fig:cnn}. The CNN is trained and its performance is evaluated with the original CTA images, the 8-bit CTA images, and the pattern spectra separately and the corresponding results are compared in the next section.

\section{Results}
\begin{figure}[h]
    \begin{subfigure}{0.47\textwidth}
        \centering
        \includegraphics[width=\textwidth,trim={0.2cm 0cm 1cm 0cm},clip]{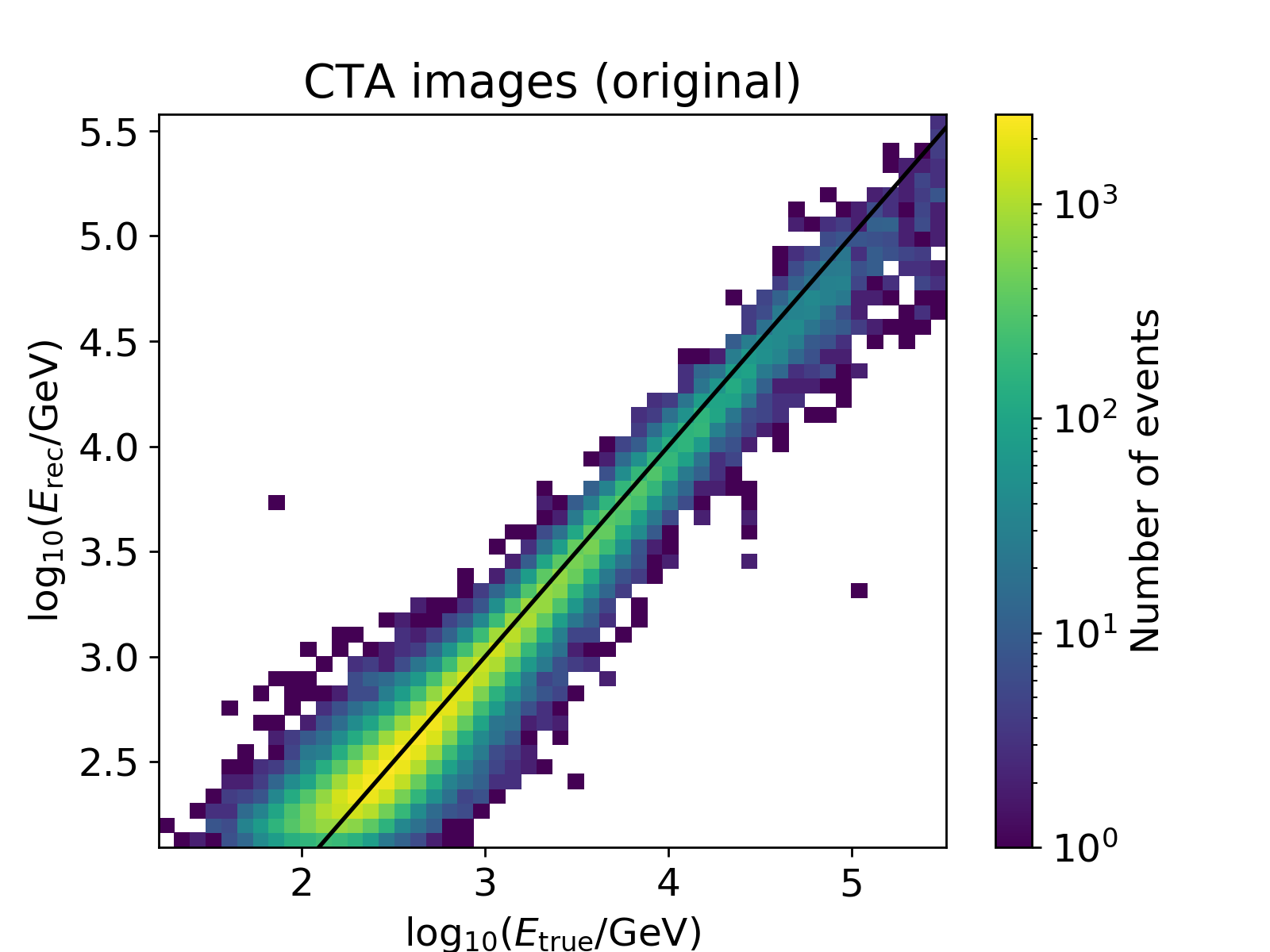}
    \end{subfigure}
    \hfill
    \begin{subfigure}{0.47\textwidth}
        \centering
        \includegraphics[width=\textwidth,trim={0.2cm 0cm 1cm 0cm},clip]{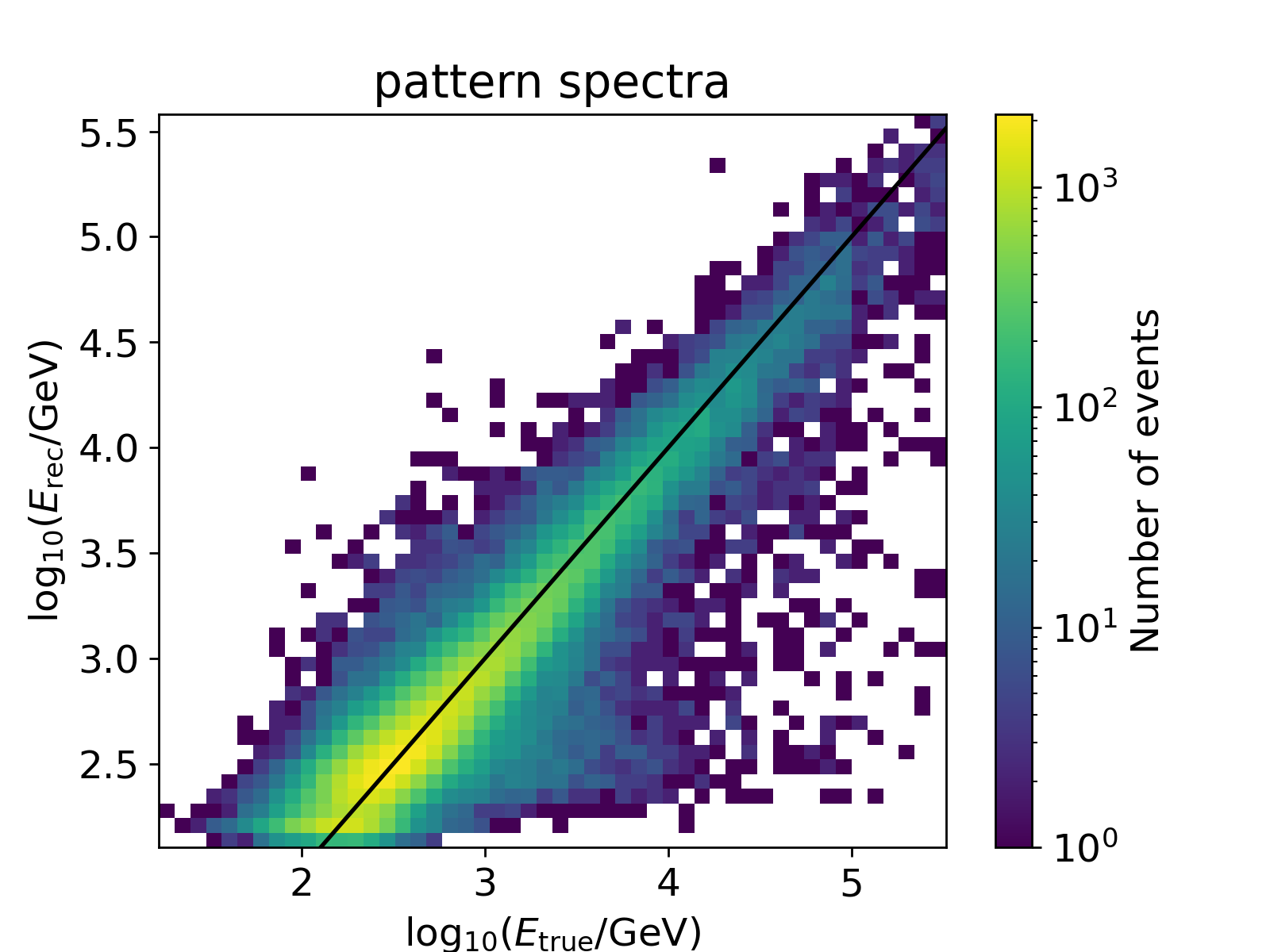}
    \end{subfigure}
	\hfill
    \begin{subfigure}{0.47\textwidth}
        \centering
        \includegraphics[width=\textwidth,trim={0.2cm 0cm 1cm 0cm},clip]{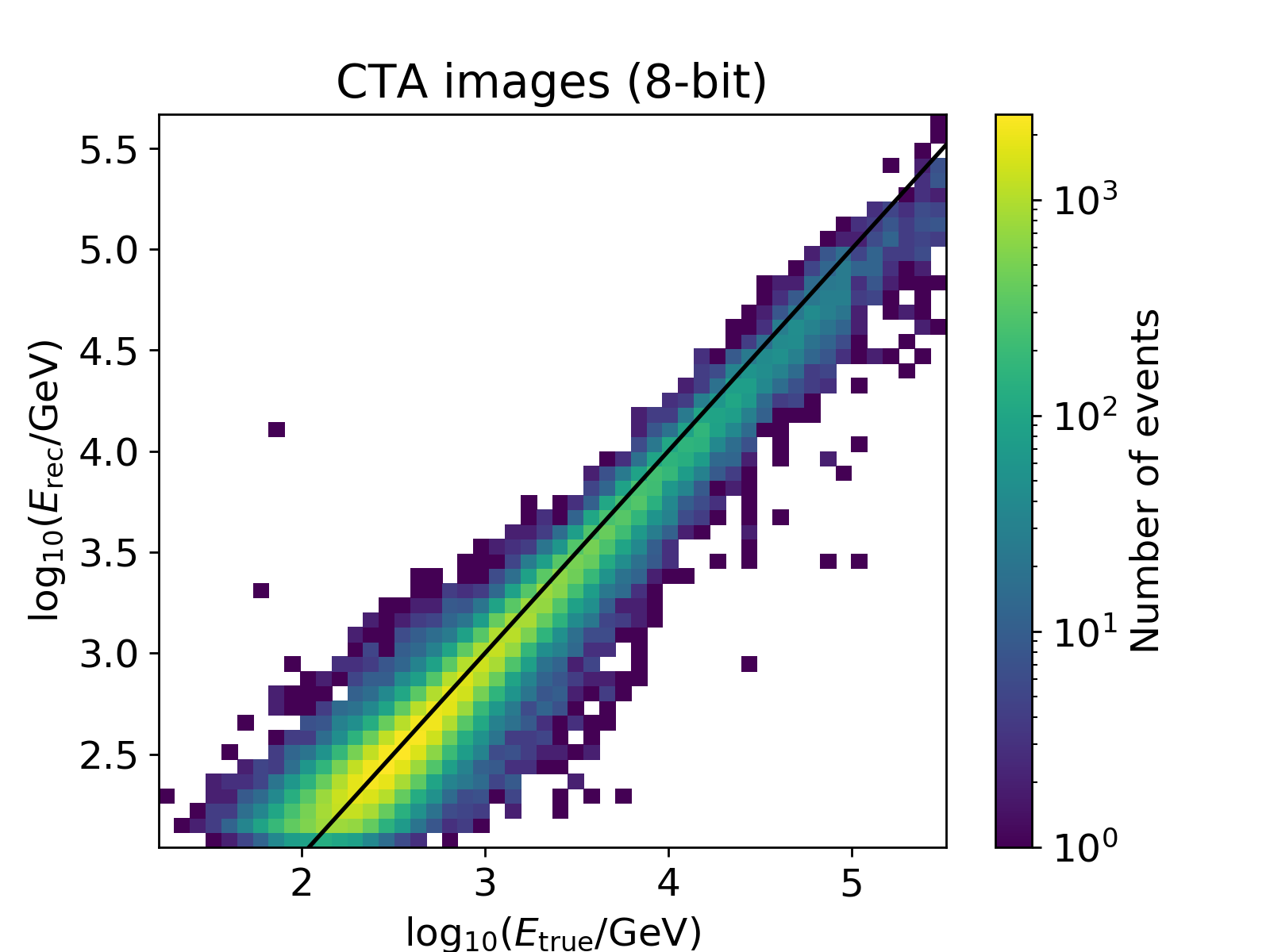}
    \end{subfigure}
    \hfill
    \begin{subfigure}{0.49\textwidth}
        \centering
        \includegraphics[width=\textwidth,trim={0cm 0.6cm 0cm 0cm},clip]{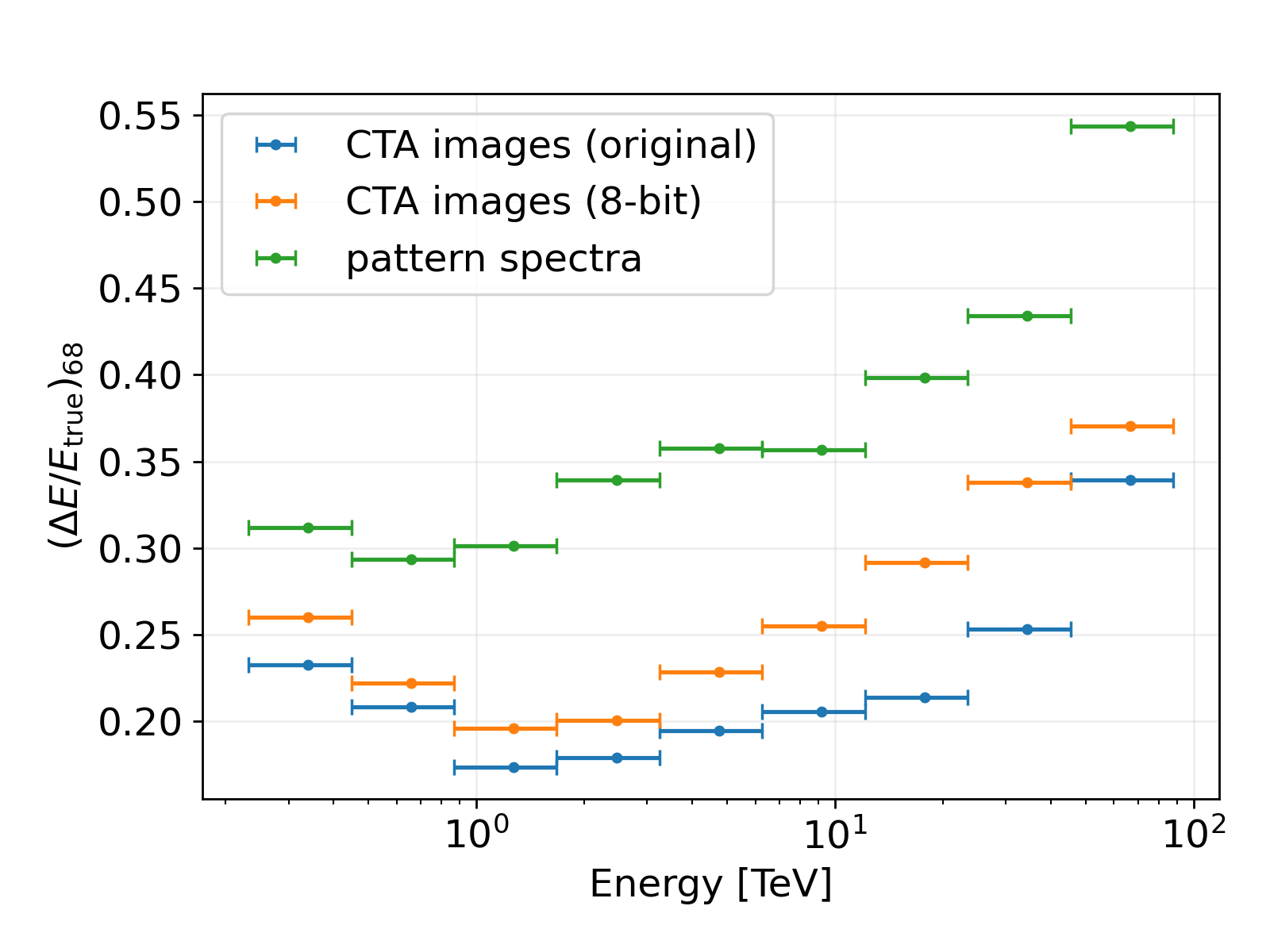}
    \end{subfigure}
       \caption{Reconstructed energy $E_{\text{rec}}$ as a function of true energy $E_{\text{true}}$ obtained with the original CTA images (top-left), 8-bit CTA images (bottom-left) and pattern spectra (top-right). Energy resolution comparison (bottom-right). The energy resolution stated in this analysis does not represent the actual energy resolution that is expected by the CTA Observatory at the end of the construction phase.}
       \label{fig:results}
\end{figure}

\noindent
The reconstructed energy $E_{\text{rec}}$ as a function of the true energy $E_{\text{true}}$ is shown in Figure~\ref{fig:results} (top-left, top-right \& bottom-left). The black line corresponds to $E_{\text{rec}} = E_{\text{true}}$. In all cases, the CNN is able to reconstruct the energy of the initial gamma ray for the majority of events. The energy scattering of the CNN with pattern spectra as input is larger compared to the results achieved with original and 8-bit CTA images. In order to quantify the results in more detail, the energy was binned logarithmically and the \textit{relative energy error} was calculated for each event via
\begin{equation}
    \dfrac{\Delta E}{E_{\text{true}}} = \dfrac{E_{\text{rec}} - E_{\text{true}}}{E_{\text{true}}}. 
\end{equation}
A histogram was created for each energy bin and the distribution was bias-corrected by subtracting the corresponding median value. The energy resolution $(\Delta E / E_{\text{true}})_{68}$ is defined as the $68^{th}$ percentile of the histogram $ | E_{\text{rec}} - E_{\text{true}} |_{\text{corr}} / E_{\text{true}}$. The comparison of the obtained energy resolution is shown in Figure~\ref{fig:results} (bottom-right). 
As already indicated in Figure~\ref{fig:results} (top-left, top-right \& bottom-left), the CNNs based on the original and 8-bit CTA images outperforms the CNN based on pattern spectra for all energies. The CNN based on 8-bit CTA images results in a lower energy resolution for almost all energies compared to the original CTA images. The energy resolution stated in this analysis does not represent the actual energy resolution that is expected by the CTA Observatory at the end of the construction phase. 

The maximum RAM and computing time needed at the \textit{Peregrine HPC cluster} on an \textit{Nvidia V100} GPU in order to train our models is shown in Table~\ref{tab:gpu}. The CNN based on pattern spectra needs $\SI{65}{\percent}$ less maximum RAM and is $\SI{41}{\percent}$ faster compared to the CNN based on the original CTA images.

\begin{table}[h]
	\centering
	\caption{Computational performance of the CNNs based on (a) original CTA images and (b) pattern spectra during training. The training was performed on a \textit{Nvidia V100} GPU at the \textit{Peregrine HPC cluster}.}
	\begin{tabular}{cccc}
			 & CTA images & pattern spectra & 1 - ratio \\ \hline
	Max. RAM & 30.26 GB   & 10.60 GB        & 65 \%     \\
	Time     & 7176 s     & 4220 s          & 41 \%    
	\end{tabular}
    \label{tab:gpu}
\end{table}

\section{Conclusions \& Outlook}
\noindent
For the first time, the energy of gamma-ray events was reconstructed by applying pattern spectra on a CNN. The fact that the pattern spectra based analysis is currently not achieving the same accuracy as the original CTA images based analysis can partly be explained by the loss of information during the conversion of the CTA images into 8-bit images before they can be put into the pattern spectra software. Thus, the pattern spectra software currently receives only information about the size and shape of the features within the image rather than getting also information about the total Cherenkov photon emission emitted by the particle shower. Since the energy of the initial gamma ray is directly proportional to the total number of Cherenkov photons emitted by the shower, this is very crucial information. However, the fact that also the CNN based on 8-bit CTA images outperforms the pattern spectra analysis for all energies might indicate that this loss of information is not the main reason for the observed difference in energy resolution. \\
The significant reduction in computational power and time needed to train our CNN indicates that pattern spectra have potential in full gamma-ray event reconstruction analyses based on CTA data. In the future, we will adjust the pattern spectra software to create pattern spectra directly from the CTA images without any loss of information. We see also a lot of room for improvement in the CNN architecture that can be adjusted more specifically on the characteristics of pattern spectra. Due to the smaller size of the pattern spectra, a simpler CNN architecture might already be sufficient to achieve a similar performance. 
Lastly, we aim to improve the background rejection of CTA by applying pattern spectra on the particle classification between gamma rays and protons.

\section*{Acknowledgments}
\noindent
This work was conducted in the context of the CTA Consortium and CTA Observatory. We gratefully acknowledge financial support from the agencies and organizations listed at \href{http://www.cta-observatory.org/consortium_acknowledgments}{http://www.cta-observatory.org/consortium$\underline{\enskip}$acknowledgments}. We would like to thank the Center for Information Technology of the University of Groningen for their support and for providing access to the Peregrine high performance computing cluster.

\clearpage

\small
\bibliography{proceedings}{}
\bibliographystyle{IEEEtran}

\clearpage
\section*{Full Authors List: The Cherenkov Telescope Array Consortium}


\scriptsize
\noindent
H.~Abdalla\textsuperscript{1}, H.~Abe\textsuperscript{2},
S.~Abe\textsuperscript{2}, A.~Abusleme\textsuperscript{3},
F.~Acero\textsuperscript{4}, A.~Acharyya\textsuperscript{5}, V.~Acín
Portella\textsuperscript{6}, K.~Ackley\textsuperscript{7},
R.~Adam\textsuperscript{8}, C.~Adams\textsuperscript{9}
% ...
\\

\noindent
1 : Centre for Space Research, North-West University, Potchefstroom, 2520, South Africa

\noindent
2 : Institute for Cosmic Ray Research, University of Tokyo, 5-1-5, Kashiwa-no-ha, Kashiwa, Chiba 277-8582, Japan

\noindent
3 : Pontificia Universidad Católica de Chile, Av. Libertador Bernardo O'Higgins 340, Santiago, Chile

\noindent
4 : AIM, CEA, CNRS, Université Paris-Saclay, Université Paris Diderot, Sorbonne Paris Cité, CEA Paris-Saclay, IRFU/DAp, Bat 709, Orme des Merisiers, 91191 Gif-sur-Yvette, France

\noindent
5 : Centre for Advanced Instrumentation, Dept. of Physics, Durham University, South Road, Durham DH1 3LE, United Kingdom

\noindent
6 : Port d'Informació Científica, Edifici D, Carrer de l'Albareda, 08193 Bellaterrra (Cerdanyola del Vallès), Spain

\noindent
7 : School of Physics and Astronomy, Monash University, Melbourne, Victoria 3800, Australia

\noindent
8 : Laboratoire Leprince-Ringuet, École Polytechnique (UMR 7638, CNRS/IN2P3, Institut Polytechnique de Paris), 91128 Palaiseau, France

\noindent
9 : Department of Physics, Columbia University, 538 West 120th Street, New York, NY 10027, USA

\noindent
% ...



\end{document}